\documentclass[aps,prb,twocolumn,groupedaddress,amsmath,amssymb]{revtex4}

\usepackage{amssymb}
\usepackage{amsmath}
\usepackage{graphicx}
\usepackage{subfigure}
\usepackage{textcomp}
\usepackage{color}
\usepackage{amsfonts}
\usepackage{bbold}
\usepackage{dsfont}
\usepackage{epsfig}
\usepackage{hyperref}

\newcommand{\arctanh}[1]{\operatorname{arctan}}

\begin{document}

\title{Predicting d$^0$ magnetism}

\author{A. Droghetti,  C.D. Pemmaraju and S. Sanvito}
\affiliation{School of Physics and CRANN, Trinity College, Dublin 2, Ireland}

\date{\today}

\begin{abstract}

Predicting magnetism originating from 2$p$ orbitals is a delicate problem, which depends on 
the subtle interplay between covalency and Hund's coupling. Calculations based on density functional 
theory and the local spin density approximation fail in two remarkably different ways. On the one hand the 
excessive delocalization of spin-polarized holes leads to half-metallic ground states and the 
expectation of room temperature ferromagnetism. On the other hand, in some cases a magnetic ground state 
may not be predicted at all. We demonstrate that a simple self-interaction correction scheme modifies 
both these situations via an enhanced localization of the holes responsible for the magnetism and possibly 
Jahn-Teller distortion. In both cases the ground state becomes insulating and the magnetic coupling 
between the impurities weak.

\end{abstract}

\maketitle

%*********************************************************************
% Introduction
%*********************************************************************

Conventional magnetism is associated with a narrow region of the periodic table, 
namely that of the 3$d$ and 5$f$ series. In these atomic shells strong Hund's coupling 
results in a high-spin configuration and therefore in the formation of localized magnetic moments. 
Then one also needs a mechanism for coupling those local moments. 
This is not universal and generally depends on the details of the compound investigated. 
Importantly the formation of magnetic moments does {\it not} imply a macroscopic magnetic state. Furthermore, 
the orbital composition of the moment is crucial for establishing the microscopic nature 
of the coupling mechanism. 

Following the criterion of having large Hund's coupling one should expect to find high-spin 
configurations and possibly magnetism also in materials with partially filled 2$p$ orbitals. This situation
takes the name of $d^0$ magnetism \cite{d0Coey}. Oxygen is the prototypical case: O$_2$ has a triplet 
ground state and solid O orders in a complex antiferromagnetic structure at low temperature \cite{AFMOx}.
Another remarkable example is that of Rb$_4$O$_6$, which was predicted to be an anionogenic 
half-metal \cite{Rb4O6HM}, and measured spin-glass \cite{Rb4O6SG}. Away from molecular solids 
however the situation is less clear. Two properties of standard solids conspire against $p$-type magnetism. 
First, the $p$ shells are usually fully filled and therefore cannot sustain a magnetic moment. Secondly 
the 2$p$ hopping integrals are usually large and so are the typical bandwidths, meaning that the chances 
of moment formation are reduced by the large kinetic energy.

Despite these complications the experimental claims for $d^0$ ferromagnetism are numerous and
include graphitic carbon \cite{Carbon} and both defective \cite{HfO2} and doped oxides \cite{Ndoping}. 
An explanation of these findings however remains controversial to date. The formation of the magnetic 
moment is usually attributed to holes localized at the defect site, either this being the molecular orbital 
associated to cation vacancies \cite{Efetov,Das} or $p$-dopants at the O sites \cite{Bouzier,Elfimov}. 
Then the magnetic coupling is justified with the degree of delocalization of the same 2$p$ shell 
responsible for the moment, in a scheme similar to the Zener model for standard magnetic 
semiconductors \cite{Dietl}. Critically most of the predictions are based on density functional 
theory (DFT) using local approximations of the exchange correlation potential (LSDA or GGA). 
These notoriously underestimate Coulomb repulsion and tend to over-delocalize the charge 
density. It is then no surprise that most of the calculations return a metallic (half-metallic) 
ground state and usually extremely large magnetic interaction. 

In this letter we demonstrate that, when strong correlation is introduced at the level of
self-interaction correction (SIC) the picture changes drastically and even the formation of a local
moment becomes a delicate issue. Consequently the mechanism for magnetic interaction can
be completely different from that described by the LSDA. We prove this idea by presenting three
prototypical cases. First we look at $F$ centers in SiO$_2$ for which LSDA predicts a delocalized
hole sustaining magnetic interaction instead of a paramagnetic self-trapped center \cite{Self,Mauri}.
Then we look at B, C and N substituting for O in MgO, for which LSDA correctly predicts a magnetic 
ground state but fails in capturing the orbital order and the insulating state. Finally we look at the case 
of Ga vacancy and dopant Zn in GaN. The former case is similar to that of $F$ centers in SiO$_2$ while 
in the latter, LSDA now fails in predicting a high-spin state.

Our calculations are performed using the standard DFT code {\sc siesta} with LSDA functional \cite{siesta}
and a development version implementing the atomic SIC scheme (ASIC) \cite{Das2}. We treat core 
electrons with norm-conserving Troullier-Martin pseudopotentials, while the valence charge density 
and all the operators are expanded over a numerical orbital basis set, including multiple-$\zeta$ and 
polarized functions \cite{siesta}. The real space grid has an equivalent cutoff larger than
500~Ry. Calculations are performed with supercells of various sizes including $k$-point
sampling over at least 10 points in the Brillouin zone. Relaxations are performed with
standard conjugate gradients until the forces are smaller then 0.04~eV/\AA. 

\begin{table}[t]
%\begin{ruledtabular}
\begin{tabular}{llccccc}\hline\hline
System & $d_\mathrm{LSDA}$  & $\mu_\mathrm{LSDA}$ &  $d_\mathrm{ASIC}$  & $\mu_\mathrm{ASIC}$  \\ \hline\hline
SiO$_2$: Al$_\mathrm{Si}$ &   1.73 (4) & 1 &  1.67 (3), 1.94 (1) & 1 \\
MgO: B$_\mathrm{O}$ & 2.19 (6) & 3 & 2.20 (6) & 3 \\
MgO: C$_\mathrm{O}$ & 2.18 (2), 2.15 (4) & 2 & 2.18 (4), 2.15 (2)  & 2 \\
MgO: N$_\mathrm{O}$ &  2.10 (6) & 1 &  2.15 (4),  2.10 (2) & 1 \\
GaN: V$_\mathrm{Ga}$ &  2.10 (4) & 3 &  2.14 (2), 2.28, 2.31 & 3 \\
GaN: Zn$_\mathrm{Ga}$ &  2.0 (4) & 0 &  1.96 (3), 2.4 (1) & 1 \\  \hline\hline
\end{tabular}
%\end{ruledtabular}
\caption{\label{Tab1} Summary table of the LSDA and ASIC calculated bond lengths, $d$ (in \AA), and 
magnetic moments, $\mu$ (in $\mu_\mathrm{B}$), for the various defects investigated in SiO$_2$, MgO 
and GaN. In bracket the number of bonds of a given length. }
\end{table}
We begin our analysis by considering the famous prototypical case of the $F$ center in SiO$_2$, i.e. Al 
substituting Si (Al$_\mathrm{Si}$). In table~\ref{Tab1} we summarize the results of our LSDA and ASIC 
calculations. These have been obtained with a 36 atom unit cell containing a single Al impurity. The LSDA 
local geometry of Al$_\mathrm{Si}$ has the four Al-O bonds of equal length, 1.73~\AA, and 
the $S$=1/2 hole spreads uniformly over the four O atoms. However relaxation with ASIC produces 
a distortion. Three of the Al-O bonds relax to 1.67~\AA, while the forth gets considerably longer (1.94~\AA). 
Such a distortion is associated with the localization of the Al-induced hole, which occupies the orbital 
along the elongated bond. This picture is consistent with the experimentally found hole self-trapping 
and demonstrates the importance of SIC in the self-trapping problem \cite{Mauri}. 

If one now calculates the magnetic interaction between two Al$_\mathrm{Si}$, a remarkable result is found.
By using a 72 atom supercell containing two Al impurities placed at second nearest 
neighbours, we have calculated the energy difference, $\Delta E$, between the ferromagnetic 
and antiferromagnetic configurations of the cell. This is about 120~meV for LSDA meaning that 
for Al concentrations around 8~\% LSDA predicts SiO$_2$:Al to be a ferromagnet with an estimated 
mean-field critical temperature of 240~K. No magnetism has ever been reported for SiO$_2$:Al at any 
concentrations and we conclude that this is simply an artifact of the excessive LSDA hole delocalization. 
When the same calculation is repeated with ASIC we find that the total energy difference drops to 1~meV, 
i.e. it is consistent with no magnetic coupling. These results set the theoretical benchmark of our 
method.
\begin{figure}[ht]
\includegraphics[width=9cm, clip=true]{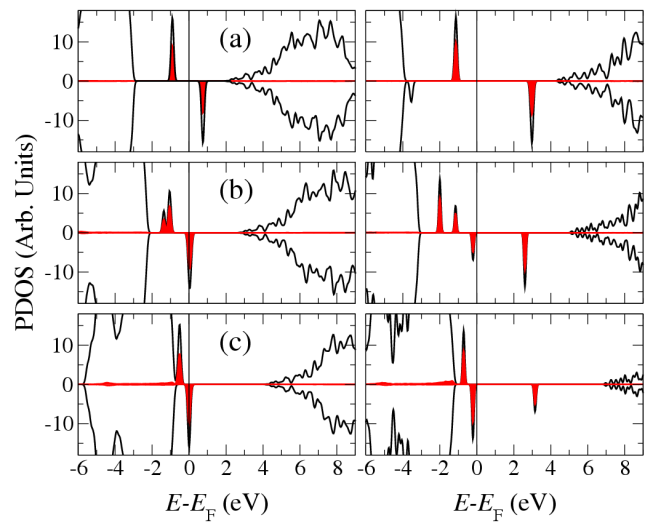}
\caption{(Color on line) Total density of states (black line) and density of states projected over the 2$p$ shells of
the anion acceptor (red line) for MgO doped respectively with B (a), C (b) and N (c). The left-hand (right-hand) 
side plots correspond to LSDA (ASIC). The majority (minority) DOS is plotted in the upper (lower) half of each 
panel.}\label{Fig1}
\end{figure}

Next we analyze the case of O substitutional defects A$_\mathrm{O}$ (A~=~B, C, N), in rock-salt MgO. These 
introduce respectively 3, 2 and 1 hole, so that the filling of the A$_\mathrm{O}$ 2$p$ shell is respectively 
1/2, 2/3 and 5/6. In figure \ref{Fig1} we report the density of states for a 96 atom supercell doped with a single 
A$_\mathrm{O}$ and projected over the A$_\mathrm{O}$ 2$p$ orbitals. For all the impurities both the LSDA 
and ASIC return a spin-polarized ground-state with a magnetic moment as large as the number of holes.
The details of the charge distribution around A$_\mathrm{O}$ are however profoundly different in LSDA and 
ASIC.

Let us look at the single acceptor N$_\mathrm{O}$ first. A magnetic ground state means that the hole is 
fully spin-polarized. The problem is then that of distributing a single hole among the three 2$p$-orbitals 
forming the chemical bonds with the Mg cations. As for SiO$_2$ also in this case LSDA predicts a spin-polarized 
ground state with the Fermi level cutting across a 2/3 filled minority N-2$p$ impurity band. LSDA atomic 
relaxation is isotropic with a N-Mg bond-length slightly larger (2.10\AA) than that between O and Mg of 
2.08\AA. This half-metallic ground state, together with the considerable energy overlap between the O 
2$p$ and N 2$p$ orbitals (see Fig.\ref{Fig1}), is suggestive of a N ferromagnetic order via an impurity 
band Zener mechanism. In fact the calculated $\Delta E$ for a 96 atom cell and second nearest neighbours 
is 120~meV and ferromagnetic. 

\begin{figure}[ht]
\includegraphics[width=7cm, clip=true]{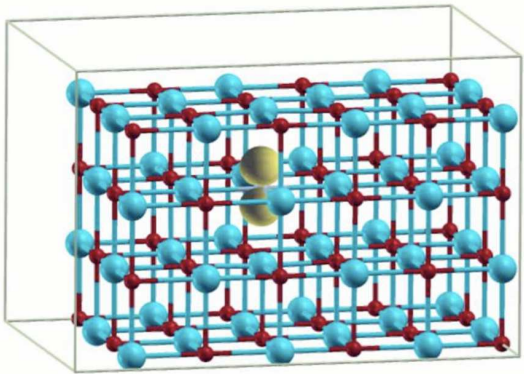}
\caption{(Color on line) Magnetization isosurfaces calculated with ASIC for N$_\mathrm{O}$ in MgO. 
Note that the $S$=1/2 hole is completely localized along the $p_y$ orbital and it is not evenly 
distributed along all the bonds. Color code: blue=Mg, red=O, N is not displayed for clarity.}\label{Fig2}
\end{figure}

Note however that this partially filled degenerate $p$ configuration is sensitive to Jahn-Teller distortion. 
This is not captured by LSDA, due to the erroneous over-delocalization of the hole, but it can be 
described by an ASIC calculation. ASIC in fact leads to the expansion of the two of six N-Mg bonds 
(2.15~\AA) with the consequent N-2$p$ levels splitting into a doubly degenerate occupied level just above the 
MgO valence band and an empty singlet. These are separated by a crystal field energy of approximately 
3~eV. In this distorted configuration the magnetic moment is entirely localized over the longer of the $p$ 
bonds as it can be clearly seen from Fig.~\ref{Fig2} where we present the ASIC magnetization isosurface.

As it is, MgO:N is an insulator and the only mechanism left for magnetic coupling between N$_\mathrm{O}$ 
is superexchange. This is expected to be extremely weak, since the first available Mg filled shells to mediate 
the virtual transition necessary for superexchange are the 2$p$, which are extremely  deep, leading practically 
to no coupling. In fact the calculated coupling at second nearest neighbour is only 1~meV. Note also that, at 
variance to the standard Jahn-Teller distortion for $d$ orbitals, the $p$-type distortion investigated 
here is frustrated, and each of the three degenerate 2$p$ orbitals can transform into the singlet. 
This introduces an additional complication to the superexchange mechanism, since only orbitals sharing 
the same angular momentum about the bonding axis can participate to the coupling. Finally 
we note that a similar Jahn-Teller distortion was predicted by Elfimov et al. \cite{saw} for N-doped SrO.

Let us now move to MgO:C. This time the minority C-2$p$ shell is singly occupied leading to a magnetic moment 
of 2~$\mu_\mathrm{B}$ for both LSDA and ASIC. For C$_\mathrm{O}$ also LSDA predicts Jahn-Teller distortion 
with four 2.15~\AA\  and two 2.18~\AA\ long Mg-C bonds. One hole distributes entirely over the 2$p$ orbital 
associated to the long bond but the second spreads evenly over
the remaining orbitals again giving a half-metallic ground state (Fig.\ref{Fig1}b). ASIC changes the Jahn-Teller 
distortion into two short (2.15~\AA) and four long (2.18~\AA) bonds, basically reversing the LSDA geometry. 
In addition ASIC correctly describes the level occupation giving a filled singlet (shorter bond) separated by 
about 2.5~eV from an empty doublet. 

The triple acceptor B$_\mathrm{O}$ presents a somewhat less critical situation. Now the B 2$p$ impurity band 
is half-filled and positioned completely within the MgO band-gap. The strong Hund's interaction produces a 
magnetic ground state with a magnetic moment of 3~$\mu_\mathrm{B}$/B for both LSDA and ASIC. For such 
an occupation the Jahn-Teller mechanism does not operate and the LSDA and ASIC bandstructures are qualitatively 
similar. 

When we look at the magnetic coupling we find for both N and C a situation very similar to that of Al in SiO$_2$,
i.e. a large magnetic coupling in LSDA and practically no coupling in ASIC. The situation is rather different for B
where even LSDA does not predict magnetic coupling. Does this mean that ferromagnetism cannot exist in these
compounds? The answer to this question is not simple and essentially relates to the behaviour of the impurities 
with additional doping. For instance by adding a fractional electronic charge to C$_\mathrm{O}$ one should 
find a situation intermediate between that on C$_\mathrm{O}$ and N$_\mathrm{O}$. Such a charge however 
feels a weaker nuclear potential and certainly localizes less. Although it is likely that it will occupy one of the 
empty C$_\mathrm{O}$ $p$ orbitals, leading once again to a half-metallic band-structure, the quantitative details 
of its exact localization are not easy to predict. Since the ferromagnetism is the results of the subtle interplay 
between localization, necessary for the magnetic moment formation, and de-localization, necessary for the 
magnetic coupling, we believe that a more accurate electronic structure method capable of taking all these 
factors into account is needed. 

We finally move our attention to Ga vacancy (V$_\mathrm{Ga}$) and of Zn substitutional for Ga 
(Zn$_\mathrm{Ga}$) in GaN. The first is again a triple acceptor and it was recently proposed as
possible source of $d^0$ ferromagnetism \cite{Dev}, while the second introduces only one hole, which is
predicted non-magnetic by LSDA. Our results, obtained with a 64 atoms supercell are summarized again in 
table \ref{Tab1}, where we can observe that LSDA and ASIC agree on the magnetic ground state of 
V$_\mathrm{Ga}$, but give respectively a non-magnetic and a magnetic $S$=1/2 state for Zn$_\mathrm{Ga}$. 
The non-magnetic state found in LSDA is resistant to supercell size and persists to cells as large as 256 atoms, 
which are already large enough to decouple the impurities located in the periodic mirror cells. 

\begin{figure}
\includegraphics[scale=0.40]{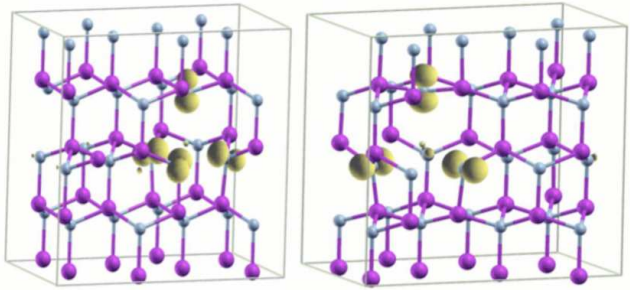}
\caption{(Color on line) Isosurface of the 3 holes (the magnetization) associated to V$_\mathrm{Ga}$ in GaN as calculated 
with LSDA (left) and ASIC (right). In LSDA the holes are distributed uniformly over all the four N ions surrounding the 
vacancy, while they localize around the longer three bonds in ASIC.}
\label{Fig3}
\end{figure}

Let us consider first the case of V$_\mathrm{Ga}$. Here the situation is somehow similar to that of 
MgO:N, where the magnetic state is reproduced already at the level of LSDA, but the spatial distribution 
of the moment changes radically when SIC is considered. In LSDA, in fact, the four N surrounding the vacancy 
move outwards so that the distance with the vacant site is 2.1\AA, about 8\% longer than that between Ga and 
N. Such displacement is isotropic and the three holes spread uniformly over the four available bond directions 
of the wurtzite lattice (see figure \ref{Fig3}). Thus the material turns out to be half metallic \cite{Dev}. The ASIC 
picture is however rather different. Also in this case the SIC allows the hole to be localized. The ASIC relaxation 
leads to two bonds of 2.14\AA~and two longer of bonds of slightly different lengths. 
The longest of the four is fully filled and the three holes localize on the 
remaining three~(Fig.~\ref{Fig3}). Such a situation reminds that of Al$_\mathrm{Si}$ and it is a further demonstration
of the effects of self-trapping. Also for GaN:V$_\mathrm{Ga}$ self-trapping massively suppresses the magnetic 
coupling energy which reduces from 145~meV (antiferromagnetic) in LSDA (for two V$_\mathrm{Ga}$ placed 
at 6.3\AA\ from each other), to 1~meV for ASIC.

The situation for the single acceptor Zn$_\mathrm{Ga}$ is dramatically different with LSDA giving a non-magnetic 
ground state with the hole evenly distributed over the four almost tetrahedrally coordinated bonds. These are 
just slightly expanded to about 2~\AA. ASIC again captures the Jahn-Teller distortion: three Zn-N bonds re-adjust 
to a distance of 1.96~\AA, while the forth expands considerably to 2.4~\AA. The hole now completely localizes 
over the longest bond, and due to the enhanced degree of localization Hund's coupling spin splits the energy level 
in forming a spin 1/2 ground state. The ASIC electronic structure then is that of Zn$_\mathrm{Ga}$ being a paramagnetic 
deep trap instead of a shallow non-magnetic acceptor.
%, as demonstrated recently experimentally.
%
\begin{figure}
\includegraphics[scale=0.39]{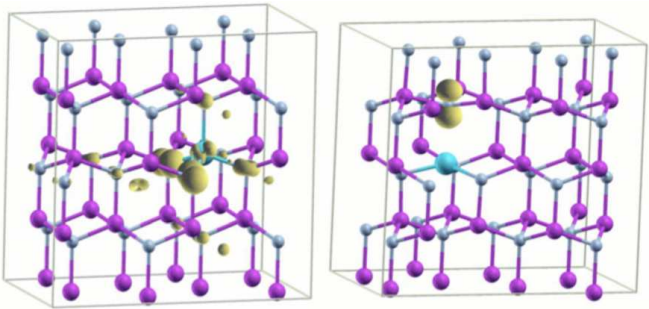}
\caption{(Color on line) Isosurface of the hole (the magnetization) associated to Zn$_\mathrm{Ga}$ in GaN as calculated 
with LSDA (left) and ASIC (right). In LSDA the hole is unpolarized and distributed uniformly over all the four N ions surrounding 
the vacancy, while it spin-splits and localizes around the longer bond in ASIC.}
\label{Fig4}
\end{figure}

In conclusion we have demonstrated that the problem of predicting the magnetic ground state of $p$-type impurities is
an extremely delicate one. In general standard LSDA is unable to capture Jahn-Teller distortion and systematically
underestimates the electron localization. As a consequence Hund's coupling is also underestimated. In the best case
scenario a magnetic ground state is still predicted, but the spatial distribution of the magnetic moment is inaccurate. This typically 
leads to a metallic (via impurity band) ground state, strong exchange coupling between the magnetic impurities, and
the expectation of room temperature ferromagnetism. The removal of the self-interaction error results in Jahn-Teller
distortion and the creation of insulating ground state with little to no magnetic coupling between the impurities. 
A more drastic situation is however encountered when the erroneous LSDA delocalization strongly suppresses Hund's 
coupling and produces a non-magnetic ground state. Also in this case ASIC is able to fix the problem and predicts
non-vanishing magnetic moment localized at distorted bonds.

%*********************************************************************
% Acknowledgment
%*********************************************************************

This work is sponsored by Science Foundation of Ireland under the grants 07/IN.1/I945 and 07/RFP/MASF238. 
Computational resources have been provided by the HEA IITAC project managed by the Trinity Center for High 
Performance Computing and by ICHEC.

\end{document}